\documentclass[useAMS,usenatbib]{mn2e}

\usepackage{amssymb}
\usepackage{graphicx}
\usepackage{hyperref}

\def\beq{\begin{eqnarray}}
\def\eeq{\end{eqnarray}}

\title[Gas clumping in galaxy clusters]{Gas clumping in galaxy clusters}

\author[D. Eckert et al.]{D. Eckert$^{1,2}$\thanks{E-mail: Dominique.Eckert@unige.ch}, M. Roncarelli$^{3,4}$, S. Ettori$^{4,5}$, S. Molendi$^{2}$, F. Vazza$^{6,4}$, 
\newauthor
F. Gastaldello$^{2,7}$, and M. Rossetti$^{8,2}$
\\
\\
$^{1}$Astronomical Observatory of the University of Geneva, ch. d'Ecogia 16, 1290 Versoix, Switzerland\\
$^{2}$INAF - IASF-Milano, Via E. Bassini 15, 20133 Milano, Italy\\
$^{3}$Universitˆ di Bologna, Dipartimento di Fisica e Astronomia, viale Berti Pichat 6/2, 40127 Bologna, Italy\\
$^{4}$INAF - Osservatorio Astronomico di Bologna, Via Ranzani 1, 40127 Bologna, Italy\\
$^{5}$INFN, Sezione di Bologna, viale Berti Pichat 6/2, 40127 Bologna, Italy\\
$^{6}$Hamburg Observatory, Gojansbergsweg  112, 21029 Hamburg, Germany\\
$^{7}$Department of Physics and Astronomy, University of California at Irvine, 4129 Frederick Reines Hall, Irvine, CA 92697-4575, USA\\
$^{8}$Universit\` a degli studi di Milano, Dip. di Fisica, via Celoria 16, 20133 Milano, Italy
}

\begin{document}

\date{Received - Accepted}

\maketitle

\begin{abstract}
The reconstruction of galaxy cluster's gas density profiles is usually performed by assuming spherical symmetry and averaging the observed X-ray emission in circular annuli. In the case of a very inhomogeneous and asymmetric gas distribution, this method has been shown to return biased results in numerical simulations because of the $n^2$ dependence of the X-ray emissivity. We propose a method to recover the true density profiles in the presence of inhomogeneities, based on the derivation of the azimuthal median of the surface brightness in concentric annuli. We demonstrate the performance of this method with numerical simulations, and apply it to a sample of 31 galaxy clusters in the redshift range 0.04-0.2 observed with \emph{ROSAT}/PSPC. The clumping factors recovered by comparing the mean and the median are mild and show a slight trend of increasing bias with radius. For $R<R_{500}$, we measure a clumping factor $\sqrt{C}<1.1$, which indicates that the thermodynamic properties and hydrostatic masses measured in this radial range are only mildly affected by this effect. Comparing our results with three sets of hydrodynamical numerical simulations, we found that non-radiative simulations significantly overestimate the level of inhomogeneities in the ICM, while the runs including cooling, star formation, and AGN feedback reproduce the observed trends closely. Our results indicate that most of the accretion of X-ray emitting gas is taking place in the diffuse, large-scale accretion patterns rather than in compact structures.
\end{abstract}

\begin{keywords}X-rays: galaxies: clusters - Galaxies: clusters: general - cosmology: large-scale structure\end{keywords}

\section{Introduction}

As the largest gravitationally-bound structures in the present Universe, galaxy clusters are concerned by accretion of smaller-scale halos and smooth gas from the large-scale structure. Still at the present epoch, the matter distribution in clusters is affected by accretion and merging processes, which causes inhomogeneities in the observed intracluster medium (ICM) and dark matter distributions. These processes are thought to be enhanced close to the outer halo boundaries, which host the transition between the virialised cluster region and the infalling material from the large-scale structure \citep[see][for a recent review]{reiprich13}. Since the X-ray emissivity of the cluster's hot gas scales like $n_e^2$, the density contrasts in X-ray images are enhanced, and inhomogeneities in the gas distribution can bias the recovered gas density profiles high \citep{mathiesen,nagai}. This will in turn bias the observed entropy \citep{nagai}, gas mass \citep{roncarelli13}, and hydrostatic mass \citep[see][and references therein]{ettori13}. 

The level of inhomogeneities in the ICM and the associated bias is usually characterised through the \emph{clumping factor} $C=\langle n_e^2\rangle/\langle n_e\rangle^2$, where $\langle\cdot \rangle$ denotes the mean inside spherical shells. The radial dependence of this quantity has been studied in detail in numerical simulations \citep{nagai,vazza12c,zhuravleva13,roncarelli13}, leading to consistent results between the various kinds of simulations. The clumping factor is found to increase steadily with radius, from negligible values in the central regions to values significantly larger than unity around $R_{200}$. Conversely, in real galaxy clusters the clumping factor and its radial dependence are still largely unknown. From the excess gas fraction observed in the \emph{Suzaku} observations of a narrow arm of the Perseus cluster, \citet{simionescu} inferred a very large value $\sqrt{C}\sim3-4$ around $R_{200}$. Instead, using a sample of 18 systems observed with \emph{ROSAT} and \emph{Planck}, \citet{eckert13a} used the deviations of the observed entropy profiles from self similarity to infer an average clumping factor of $\sqrt{C}\sim1.2$ around $R_{200}$, in better agreement with the predictions of numerical simulations \citep[see also][]{walker13,urban13}. These methods are however indirect and rely on several strong assumptions. A direct measurement of the actual level of gas clumping is still lacking.

Recently, \citet{morandi13} suggested to use the standard deviation of the surface brightness fluctuations in concentric annuli as a tracer of the gas clumping factor \citep[see also][]{churazov12}, and applied this method to \emph{Chandra} observations of A1835, for which they measured $\sqrt{C}\sim1.5$ at $R_{200}$. This method is promising, but requires very high quality data. A similar approach was adopted in numerical simulations by \citet{roncarelli13}, who related the azimuthal scatter in narrow sectors \citep{vazzascat,e12} to the level of asymmetries in numerical simulations. The azimuthal scatter was found to trace closely the large-scale asymmetries in clusters, but fails at determining the bias introduced by small-scale structures ($\lesssim 100$ kpc). 

In this paper, we present a method to recover unbiased density profiles based on the azimuthal median of the surface brightness in each radial annulus. The azimuthal median allows us to alleviate significantly the effect of the emissivity bias, since it is robust against strong outliers. Based on both hydrodynamical and synthetic numerical simulations, we demonstrate that the azimuthal median is a good tracer of the true cluster surface-brightness profile out to large radii (see Sect. \ref{sec:validation}). In Sect. \ref{sec:application}, we present an efficient algorithm to recover the azimuthal median from X-ray observations. This algorithm is applied to a sample of 31 clusters with available \emph{ROSAT}/PSPC pointed observations in Sect. \ref{sec:results}. Finally, we present a comparison of our observational results with three different sets of numerical simulations and discuss our results in Sect. \ref{sec:discussion} and conclude our paper in Sect. \ref{sec:conclusion}.

\section{Simulated datasets}
\label{sec:sims}

Here we briefly describe the numerical simulations used throughout the paper and the simulated cluster samples drawn from these simulations.

\subsection{GADGET simulations}
\label{sec:gadget}

We used simulated clusters from a set of 62 clusters and groups simulated with the Tree smoothed particle hydrodynamics (Tree-SPH) code \texttt{GADGET-3} \citep{roncarelli13,bonafede11,planelles13}. These clusters are drawn from a large dark-matter-only simulation with a periodic box of 1$h^{-1}$ Gpc on a side. The regions around $7R_{\rm vir}$ from the centre of the 24 most massive objects (all with $M_{\rm vir} > 10^{15}h^{-1} M_\odot$) were identified and then re-simulated at higher resolution using the zoomed initial conditions technique \citep{tormen97}, thus including as well the lower-mass halos at the vicinity of these massive systems. An additional set of 5 clusters with masses $M_{200} = 1-5\times10^{14} h^{-1}M_\odot$ was selected to add statistics at the lower mass scale and re-simulated with the same method, reaching a total of 29 re-simulated regions. The final sample of re-simulated systems comprises 28 clusters and 34 groups, where the boundary between groups and clusters is set to a mass threshold of $M_{200}=5\times10^{14}h^{-1}M_\odot$. Roughly half of the systems (30 out of 62) are classified as relaxed, the remaining 32 systems being dynamically active. 

To investigate the impact of additional physics on our results, we used two different re-simulation runs, one including only gravitational and hydrodynamical effects (hereafter referred to as \emph{non-radiative}) and another including gas cooling, star formation, galactic winds, and AGN feedback \citep[hereafter called \emph{CSF+AGN}]{planelles13}. Radial profiles of gas density and clumping factor were extracted from both runs to be compared with observations (see Sect. \ref{sec:discussion}). In addition, we also used ``residual'' profiles, which were obtained by masking the densest regions to disentangle the effect of small-scale clumping from that of smooth accretion patterns \citep{roncarelli13}. In more detail, in each radial shell the SPH particles are sorted  according to their density. Then, the regions comprising the densest 1\% of the total volume of the radial bin are identified as clumps and excised for the computation of the density and clumping factor profiles, which allows us to retain only large-scale accretion patterns. For more details on the method and the dataset, we refer the reader to \citet{roncarelli13}.

\subsection{ENZO simulations}
\label{sec:enzo}

This dataset consists in a set of 20 clusters with virial masses in the range $6 \times 10^{14} \leq M/M_{\odot} \leq 3 \times 10^{15}$ extracted from a total cosmic volume of $L_{\rm box} \approx 480$ Mpc/h. These clusters were simulated with the grid code ENZO using adaptive mesh refinement tailored to probe down to a resolution of $\approx 25 ~\rm kpc/h$ for most of the virial volume, neglecting the effects of radiative cooling.  The dynamical state of each object were assigned following in detail its matter accretion history for $z \leq 1.0$. According to this definition, at $z=0$ the sample contains ten post-merger systems (i.e. clusters with a merger with a mass ratio larger than $1/3$ for $z \leq 1$), six merging clusters (i.e. systems in which there is evidence of an ongoing large-scale merger, but the cores of the colliding systems have not yet crossed each other) and four relaxed clusters. We refer the reader to \citet{va10kp} for the full list of cluster parameters (e.g., gas and total mass, virial radius) and to \citet{e12,eckert13a} for previous comparisons between observations and simulations performed with the same sample.

\section{The azimuthal median method}
\label{sec:method}

\begin{figure}
\resizebox{\hsize}{!}{\includegraphics{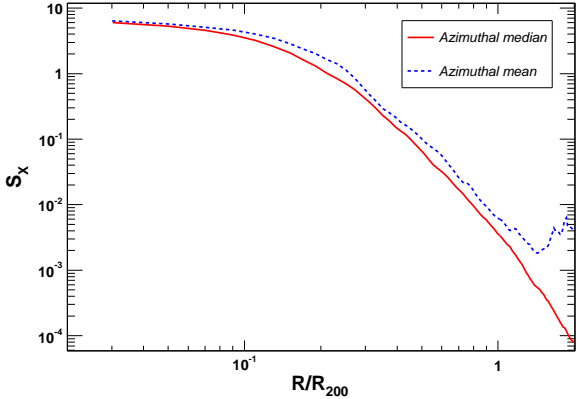}}
\caption{Average surface-brightness profiles (in arbitrary units) for the sample of simulated clusters of \citet{vazza11a} obtained by performing the azimuthal mean (dashed blue) and the azimuthal median (solid red). \label{fig:meanvsmedian}}
\end{figure}

\subsection{Method}
\label{sec:validation}

In a recent paper, \citet{zhuravleva13} used a set of 16 clusters simulated using the Adaptive Refinement Tree (ART) code \citep{nagai07} to study the distribution of gas density in concentric shells. They showed that the gas density distribution at a given radius from the cluster center follows approximately a log-normal distribution, plus a skewed tail towards high densities. Even though it is made of a small number of cells in the simulations, this high-density tail biases the observed mean density profiles significantly in mock X-ray observations because of the $n_e^2$ dependence of the X-ray emissivity. Conversely, \citet{zhuravleva13} concluded that the median of the distribution coincides with the peak of the log-normal distribution, and thus it is a good tracer of the typical cluster behaviour in the presence of inhomogeneities. This conclusion appears to hold as well on the observational side (see Appendix \ref{app:sbdist}).

Observationally, one only has access to the projected 2D surface brightness distribution, and the 3D density profile is typically inferred by deprojecting the projected surface brightness profile, which is obtained by taking the mean surface brightness in each radial bin, under the assumption of spherical symmetry. Since the median is robust against outliers, we propose to use the azimuthal median of the surface brightness as an unbiased estimator of the true surface brightness. In principle, assuming that the median surface brightness is unbiased, the bias induced by the X-ray emissivity, hereafter the \emph{emissivity bias} $b_X$, in a given annulus could then be estimated by taking the ratio between the mean and the median surface brightness,

\begin{equation} b_X= \frac{S_{X, \rm mean}}{S_{X, \rm median}}. \label{eq:clf}\label{eq:bx}\end{equation}

To validate this hypothesis, we applied this technique to two different sets of numerical simulations. In the first case, we used projected $L_X$ images drawn from ENZO sample described in Sect. \ref{sec:enzo}. In the second, we performed synthetic simulations by assuming a 3D density distribution given by a beta model and adding inhomogeneities of several kinds: i) a filament of gas extending in the N-S direction; ii) a triaxial underlying halo shape; iii) a number of gas clumps with a radius of 20 kpc. X-ray surface-brightness maps were then created by projecting the resulting 3D emissivity along the line of sight. 

\begin{figure}
\resizebox{\hsize}{!}{\includegraphics{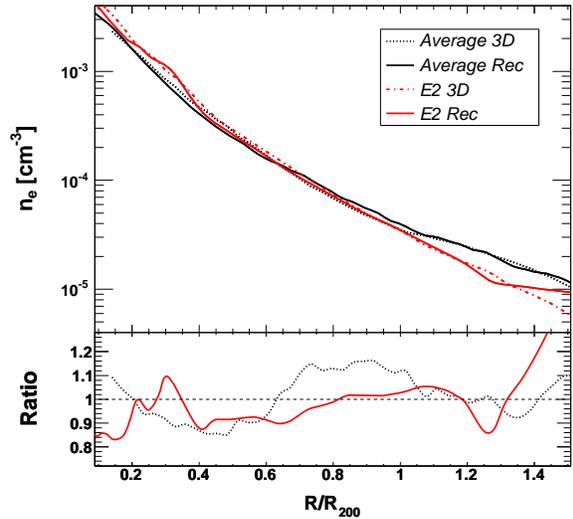}}
\caption{Comparison between the true 3D density profiles (dashed lines) and the profiles reconstructed with our method from projected $L_X$ images (solid lines) for the sample of simulated clusters of \citet{vazza11a}. The black lines show the result for the average of the 20 clusters, while the red lines show the result for one very perturbed system, E2. The bottom panel shows the ratio between the reconstructed profiles and the 3D ones for the entire sample (dashed black) and the E2 cluster (solid red). \label{fig:comparison}}
\end{figure}

Both for the hydrodynamical and the synthetic simulations, a blind analysis was performed on the simulated images. First, surface brightness profiles were extracted by computing the azimuthal median in concentric annuli. The profiles were then deprojected using a non-parametric onion-peeling technique \citep{kriss,mclaughlin99} and converted into radial density profiles. For the synthetic simulations, we found a very good agreement between the true and the reconstructed density profiles, in all of the situations considered here. Even when adding as many as $10^3$ randomly-distributed clumps in the simulated image, our method is able to recover the underlying density profile with an accuracy of a few percent at all radii. The addition of significant triaxiality to the halo did not affect our results \citep[in agreement with the findings of][]{buote12}, and similar conclusions were reached for the inclusion of a filamentary structure. In the case of the hydrodynamical simulations, we show in Fig. \ref{fig:meanvsmedian} the average surface-brightness profiles (in arbitrary units) obtained using our azimuthal median method and with the standard azimuthal mean. We can see that the azimuthal median results in a much more regular profile, showing the expected smooth radial decline. In Fig. \ref{fig:comparison} we show the comparison between the average true 3D density profile in the sample of \citet{vazza11a} and the average profile obtained with our method through deprojection. On the same figure, we also show the same comparison in the case of one particularly perturbed cluster, E2 \citep[see][]{vazza11a}. As shown in the bottom panel, the deviations are typically at the level of 10\% or less, even in the case of the very perturbed system E2, and do not show any particular trend with radius. A similar result is found for the integrated gas mass. This demonstrates the ability of our method to recover the true density profiles from X-ray observations in the presence of inhomogeneities.

\subsection{Clumping factor}
\label{sec:bx_vs_C}

We remark that our definition of the emissivity bias (Eq. \ref{eq:clf}) is related, but not strictly equivalent to the usual definition of the clumping factor \citep{mathiesen}, since the emissivity bias is obtained from projected 2D data, whereas the clumping factor in numerical simulations is computed using the full 3D density information. Since the projected distance of substructures from the cluster center is always smaller than the true distance, and some obvious substructures may be invisible in projection, we do not expect a one-to-one correlation between our definition and the true 3D clumping factor. Since, as demonstrated in Fig. \ref{fig:comparison}, the deprojected azimuthal median is a good tracer of the true 3D density, the clumping factor can be formally recovered as the ratio between the deprojected mean and median profiles,
\begin{equation} C=\frac{\mbox{deproj}(S_{X, \rm mean})}{\mbox{deproj}(S_{X, \rm median})}.\label{eq:Cdep}\end{equation}

In practice, this relation is difficult to to apply to individual systems, since in the outer regions the geometry of the systems deviates significantly from spherical symmetry. It should however hold when applied to sizable cluster samples. To validate this hypothesis, we used simulated clusters from the \texttt{GADGET-3} sample (see Sect. \ref{sec:gadget}). We excluded the systems showing too many substructures in the outskirts (i.e. for which the deprojection is made impossible) from the simulated sample, keeping a subsample of 38 systems for this particular analysis. We extracted projected $L_X$ images assuming that the emissivity scales as $n^2T^{1/2}$ and excluding the gas below $10^6$ K which would be unobservable in X-rays. Then based on the projected images we extracted mean and median surface-brightness profiles as explained above, and used Eq. \ref{eq:Cdep} to estimate the 3D clumping factor.

\begin{figure}
\resizebox{\hsize}{!}{\includegraphics{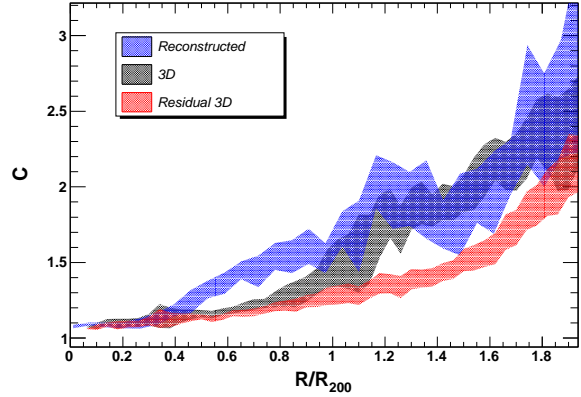}}
\caption{\label{fig:Crec}Comparison between the 3D clumping factor (black shaded area) and the clumping factor reconstructed using Eq. \ref{eq:Cdep} in the \texttt{GADGET-3} sample \citep[blue,][]{roncarelli13,bonafede11}. The red shaded area shows the 3D ``residual clumping'' obtained by cutting out the densest gas cells \citep[see][]{roncarelli13}.}
\end{figure}

In Fig. \ref{fig:Crec} we show the comparison between the average clumping factor profiles obtained using Eq. \ref{eq:Cdep} (blue) and the true 3D clumping factor profile (black). We also show the ``residual clumping factor'' obtained by masking the densest regions \citep[see Sect. \ref{sec:gadget} and][]{roncarelli13}.  In all cases, the shaded areas show the median value and standard deviation obtained using $10^4$ bootstrap samplings of the population. We can see that our method is doing a good job at reproducing the trend of the 3D gas clumping profiles in our sample of simulated systems. The clumping factor profiles recovered in this way trace more closely the total clumping than the residual one. Since the residual profile excludes the contribution of compact structures to the clumping factor, this indicates that our method is able to recover both the large-scale and small-scale inhomogeneities. Indeed, we expect that both clumps and large-scale inhomogeneities contribute to some extent to the profiles obtained using the azimuthal mean, while the azimuthal median is robust against both types of fluctuations. Thus, our method is able to recover not only the large-scale density fluctuations, but also inhomogeneities on smaller scales. 

In addition, we also investigated how projection effects impact the recovered quantities. In Appendix \ref{app:lognormal} we explore analytically the relation between emissivity bias $b_X$ and clumping factor $C$ in the log-normal formalism.

\subsection{Application to X-ray observations}
\label{sec:application}

\begin{figure}
\resizebox{1.1\hsize}{!}{\includegraphics{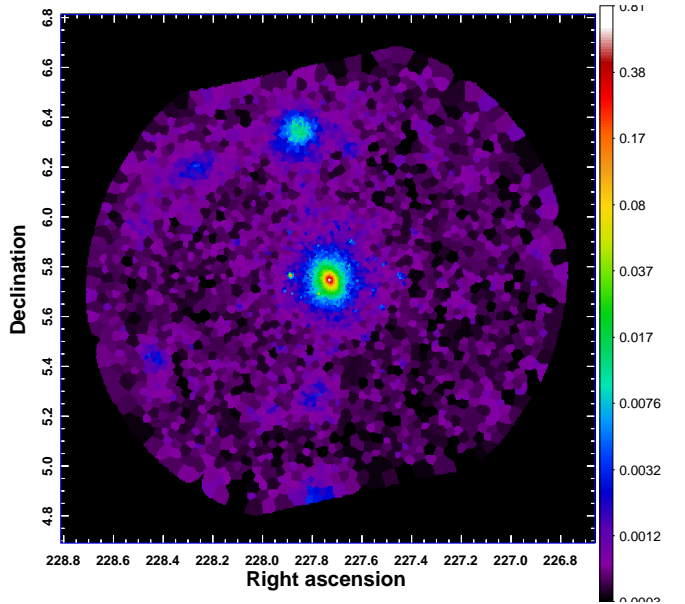}}
\caption{\emph{ROSAT}/PSPC surface-brightness image of A2029 obtained using Voronoi tessellation \citep{cappellari03}, in units of cts s$^{-1}$ arcmin$^{-2}$. Each Voronoi bin includes an average of 20 counts.\label{fig:voronoi}}
\end{figure}

We have shown in Sect. \ref{sec:validation} that in an ideal case the azimuthal median allows us to recover the density profile in an unbiased way. In practice, the computation of the azimuthal median in actual data is not straightforward, since the surface brightness in each individual pixel is low and is often affected by small-number statistics. To alleviate this effect, we make use of a 2D binning algorithm based on Voronoi tessellation \citep{cappellari03,diehl06} and rebin the images to ensure an average of 20 counts per bin to apply the Gaussian approximation. An example of binned image is shown in Fig. \ref{fig:voronoi} applied to a \emph{ROSAT}/PSPC pointed observation of A2029. For the computation of the median, we stress that requiring a minimum signal-to-noise ratio is not necessary, since we only require the error distribution to be well approximated by a Gaussian. Thus, the raw image is binned prior to background subtraction. 

From the binned image, we compute the surface brightness and its error in each bin and assign it to the included pixels. We assume that the error on each pixel of a given bin is the same and is normally distributed, such that the error on each pixel can be approximated by the error on the bin multiplied by the square root of the number of pixels in the bin,

\begin{equation} \sigma_{\rm pix} \sim \sigma_{\rm bin}\sqrt{N_{\rm pix}}, \end{equation}

where $\sigma_{\rm bin}$ follows Poisson statistics. Once the binned surface-brightness image is extracted, the surface brightness of a given annulus is simply estimated by taking the median surface brightness in the corresponding pixels. The error on the median is estimated through Monte Carlo by generating $10^4$ random realisations of the azimuthal median and computing the $1\sigma$ confidence intervals from the distribution of the different realisations. A bias-corrected surface brightness profile is then computed by repeating this procedure in each radial bin. To validate this approach, we simulated a surface brightness distribution given by a beta model and compared the confidence intervals on the median obtained using this technique with the expected results. This procedure was found to provide a good approximation of the true error bars.

We note that the ability of this method to recover the emissivity bias is limited by statistics and by the angular resolution of the instrument considered. Indeed, density variations on scales much smaller than the typical size of the Voronoi bins cannot be recovered using this method if they are uniformly distributed. It is expected from numerical simulations \citep[e.g.,][]{vazza12c} that gas clumps tend to cluster along preferential directions of accretion, such as large-scale filaments. If this is the case, inhomogeneities on scales smaller than the bin resolution would occupy a small fraction of the total volume, and such clumps would not bias the azimuthal median. However, in the general case our method is able to recover the effect of inhomogeneities only for scales larger than the resolution of the Voronoi bins.

\subsection{Cluster sample and data analysis}

We applied the method proposed here to the sample of \citet[see Table 3]{e12}. The sample is composed of 31 clusters in the redshift range 0.04-0.2 with high-quality \emph{ROSAT}/PSPC pointed observations. Among these 31 systems, 14 systems are classified as relaxed, cool-core (CC) based on their central entropy ($K_0<30$ keV cm$^2$), whereas the remaining 17 systems are classified as dynamically-active, non-cool-core (NCC, $K_0>50$ keV cm$^2$). Thanks to the large field of view and low background of the PSPC, this data is suitable for the study of the gas distribution out to large radii ($\sim2R_{500}$). In the cases where multiple PSPC observations were available for a single source, mosaic images were created and analysed. Data reduction was performed using the Extended Source Analysis Software \citep[ESAS,][]{esas}. Point sources were detected using the ESAS tool \texttt{detect} down to a fixed count rate threshold of 0.003 cps in the \emph{ROSAT} R3-7 band (0.42-2.01 keV), which corresponds roughly to a flux threshold of $3\times10^{-14}$ ergs cm$^{-2}$ s$^{-1}$. For the details of the data reduction and background subtraction procedures, we refer the reader to \citet{eckertpks,e12}. 

\begin{figure}
\resizebox{\hsize}{!}{\includegraphics{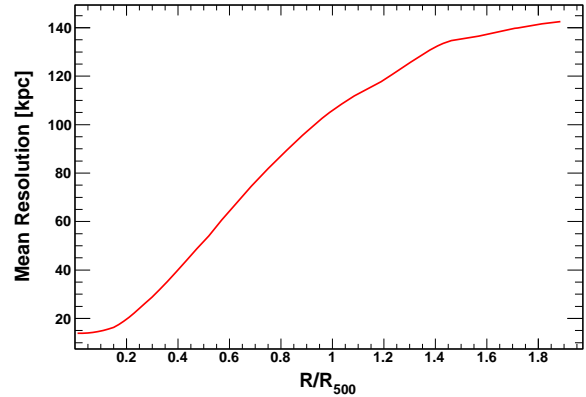}}
\caption{Mean resolution of our binned images (in kpc) as a function of radius from the cluster center.\label{fig:resolution}}
\end{figure}

Once the data were reduced, Voronoi tessellation was applied to the count maps, and vignetting correction and particle background subtraction were performed to obtain binned surface-brightness maps (see Sect. \ref{sec:application} and Fig. \ref{fig:voronoi}).Since the reliability of our method is limited by the resolution of our binned images, we computed the typical bin size as a function of distance from the cluster center in each case. In Fig. \ref{fig:resolution} we show the mean resolution of the binned images as a function of radius. At $R_{500}$, the mean resolution in our sample is roughly 100 kpc, with significant differences from one system to another. In the best cases (A1795, A3558), we reach a resolution of $\sim30$ kpc, while in the worst cases (A3158, A2163) the resolution is of the order of 200 kpc. 

From the binned images, the median surface brightness profile was then calculated and compared with the profile obtained in a standard way by averaging the surface brightness in concentric annuli. The bias-corrected surface brightness profiles were also used to extract deprojected density profiles. 

From the individual profiles, we applied a self-similar scaling \citep[see][for details]{arnaud02} by converting the surface brightness into emission measure and rescaling it by the quantity

\begin{equation}\Delta_{SSC}=\Delta_z^{3/2}(1+z)^{9/2}\left(\frac{kT}{10\mbox{ keV}}\right)^{1/2}.\end{equation}

\noindent We then obtained a median scaled emission-measure profile by interpolating the various profiles onto a common grid and taking the median value at each radius. The same procedure was applied to the profiles obtained in a standard way, and average emissivity bias profiles were then obtained using Eq. \ref{eq:clf} by comparing the average emission-measure profiles obtained with the two techniques.

\subsection{Systematic uncertainties}

Since the X-ray surface brightness in cluster outskirts is well below the background level, systematic uncertainties in the background subtraction must be properly accounted. \citet{e12} analysed a number of blank fields observed with \emph{ROSAT}/PSPC and studied the variations of the surface brightness around the mean value throughout the field of view to estimate the level of systematics. While the average surface-brightness profile of the blank fields was found to be well-matched by a constant, an excess scatter of 6\% of the average background value was measured with respect to the dispersion induced by statistics only, which \citet{e12} adopted as a systematic uncertainty in the background subtraction procedure. This value takes into account both the systematic uncertainties in the instrument calibration and the variance induced by residual point sources. We used the same value for the systematic uncertainties and added it in quadrature to all the profiles considered here.

\section{Results}
\label{sec:results}

\subsection{Average profiles}

\begin{figure}
\resizebox{\hsize}{!}{\includegraphics{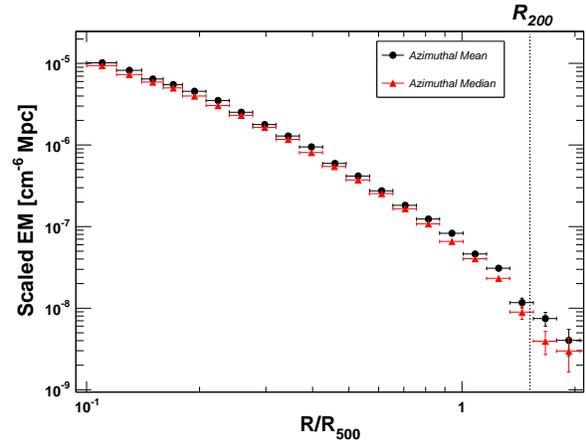}}
\caption{Median self-similar scaled emission-measure profiles obtained using the standard way (black circles) and by applying the azimuthal median technique presented here (red triangles).\label{fig:scem}}
\end{figure}

\begin{figure*}
\resizebox{\hsize}{!}{\hbox{\includegraphics{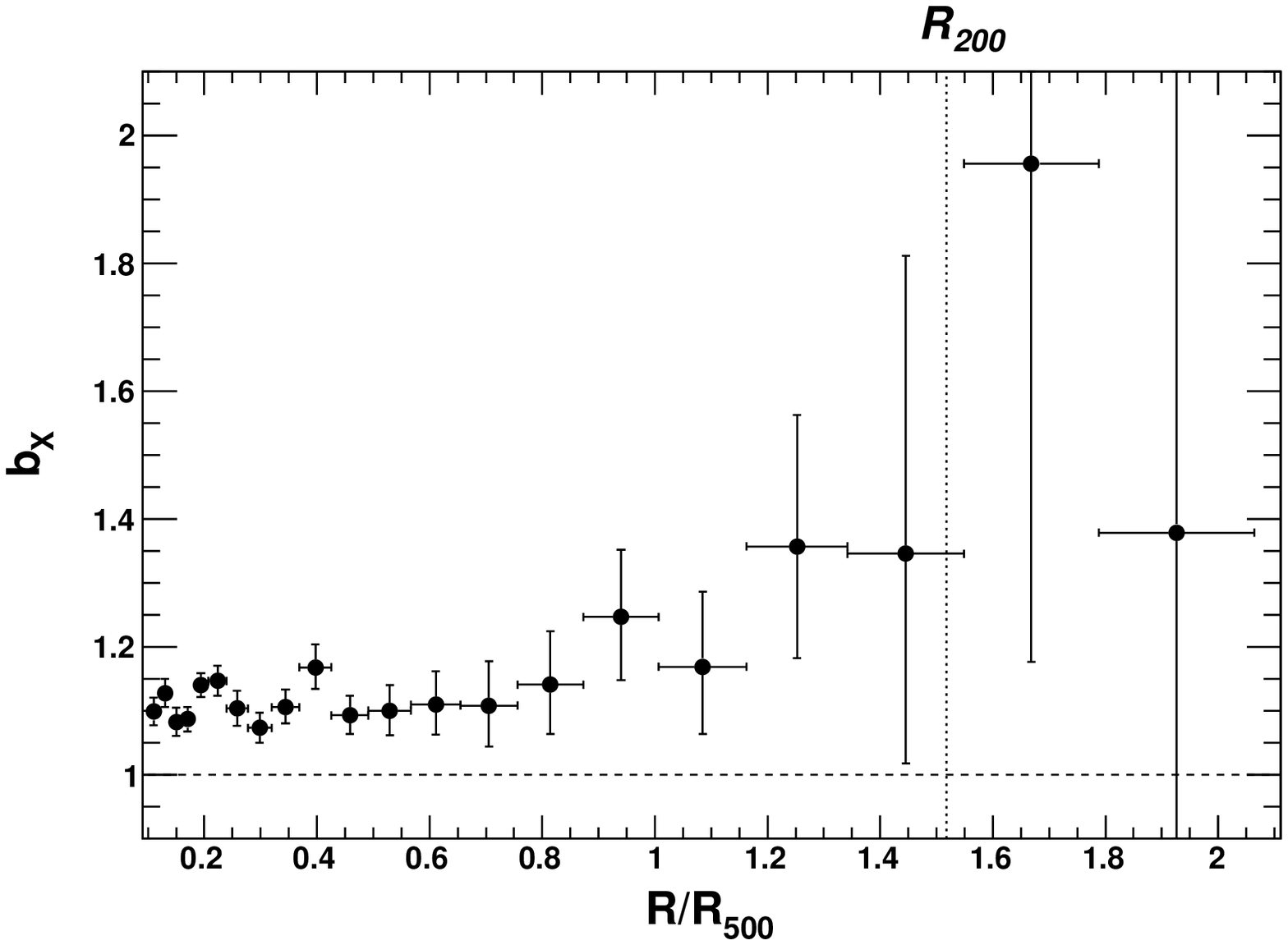}\includegraphics{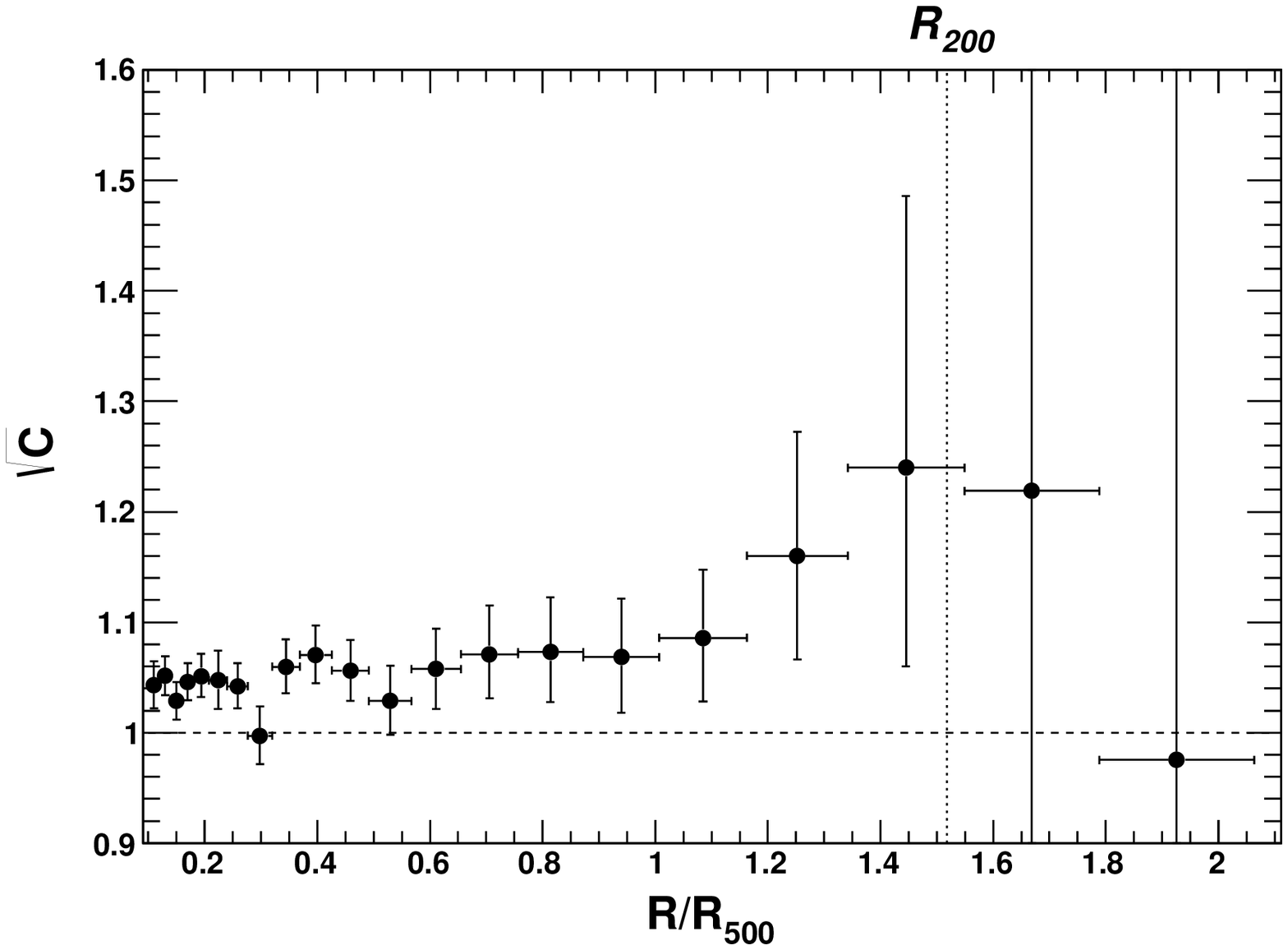}}}
\caption{\label{fig:clfprof}Average emissivity bias $b_X$ (left, see Eq. \ref{eq:clf}) and clumping factor $\sqrt{C}$ (right, see Eq. \ref{eq:Cdep}) as a function of distance from the cluster center for our sample of 31 clusters.} 
\end{figure*}

In Fig. \ref{fig:scem} we show the median scaled emission-measure profiles obtained using the standard way and by taking the azimuthal median. The differences between the two methods are mild, but significant. We note that the systematic uncertainties are negligible out to $\sim1.3R_{500}$, but dominate the error budget at $R_{200}$ and beyond, since in these regions the signal is only a fraction of the background level. The average emissivity bias $b_X$ obtained from these two profiles using Eq. \ref{eq:bx} is shown in the left panel of Fig. \ref{fig:clfprof}. We find a mild trend of increasing bias with radius, from low values $\sim1.1$ in the inner regions to slightly higher values $\sim1.5$ in the outskirts, although the error bars in this regime are large. To quantify the radial dependence of the emissivity bias, we fitted the observed emissivity bias profile with a second-order polynomial, $b_{X}=b_0+b_1(R/R_{500})+b_2(R/R_{500})^2$. We obtained values of $b_{0}=1.12\pm0.02$, $b_1=-0.10\pm0.12$, and $b_2=0.17\pm0.12$ for the best-fit parameters. We also stacked the deprojected mean and median emission-measure profiles to estimate the average clumping factor profile (see Sect. \ref{sec:bx_vs_C}). The resulting profile is shown in the right panel of Fig. \ref{fig:clfprof}. At $R_{500}$, we estimate an average clumping factor $\sqrt{C}=1.07\pm0.05$, while at $R_{200}$ we obtain $\sqrt{C}=1.25_{-0.21}^{+0.31}$.  Fitting the observed clumping factor profile with a second-order polynomial, $\sqrt{C}=A_0+A_1(R/R_{500})+A_2(R/R_{500})^2$, we measure $A_0=1.05\pm0.02$, $A_1=-0.03\pm0.09$, and $A_2=0.08\pm0.08$.

We also investigated the behaviour of the emissivity bias by splitting our sample into CC and NCC systems. In \citet{e12}, we noticed that the density profiles of NCC systems appear to systematically exceed those of CC clusters by $\sim15\%$ beyond $0.3R_{500}$ \citep[see also][]{maughan11}. As predicted by numerical simulations \citep{nagai}, a possible explanation for this difference would be a larger clumping factor in unrelaxed systems with respect to relaxed clusters, which would be caused by the presence of a larger number of substructures in the former population. Out to $R_{500}$,  we indeed find a slightly higher clumping factor in NCC clusters (with a mean value of $\sqrt{C_{\rm NCC}}=1.06\pm0.01$ in the $[0.1-1]R_{500}$ range) compared to CC systems ($\sqrt{C_{\rm CC}}=1.04\pm0.01$ in this range). These differences are however very modest and are insufficient to explain the differences between the CC and NCC cluster populations. Beyond $R_{500}$, no significant difference between the two populations can be found because of insufficient statistics.

\subsection{Bias on derived quantities}
\label{sec:derived}

Since the (squared) density is the primary observable of the ICM from X-ray observations, all the quantities derived from it are affected by the clumping factor discussed in this paper \citep[see also][]{roncarelli13}. Here we discuss the typical biases introduced in some relevant quantities by neglecting the effect of clumping.

\begin{itemize}
\item
\emph{Gas mass:} The gas mass is derived simply by integrating the deprojected density over the volume. It is a low-scatter proxy of the total mass \citep[e.g.,][]{okabe10,mahdavi13}, and could be used to estimate cluster masses in wide X-ray surveys like \emph{eROSITA}. Thus, an accurate knowledge of the effect of gas clumping on X-ray measurements of the gas mass is important. The bias in gas mass can be inferred from our data through the relation

\begin{equation}\frac{M_{\rm gas,obs}}{M_{\rm gas, true}}(r)=\frac{\int_0^r\sqrt{C}(r^{\prime})\rho(r^{\prime})\, r^{\prime2}\, dr^{\prime} }{\int_0^r \rho(r^{\prime}) \, r^{\prime2}\, dr^{\prime}}.\label{eq:mgas}\end{equation}

In Fig. \ref{fig:mgas} we show the average bias in gas mass as a function of radius for the \emph{ROSAT} sample. As expected, the bias increases steadily with radius. At $R_{500}$ we measure $\frac{M_{\rm gas,obs}}{M_{\rm gas, true}}=1.06\pm0.02$, while at $R_{200}$ we find an average bias of $1.11\pm0.04$. Fitting these profiles with a second-order polynomial, $M_{\rm gas, obs}/M_{\rm gas, true}=B_0+B_1(R/R_{500})+B_2(R/R_{500})^2$, we obtain $B_0=1.04\pm0.01$, $B_1=-0.02\pm0.03$, and $B_2=0.04\pm0.03$. The recovered biases are small, but significant, and could influence the scaling relation between gas mass and total mass. Differences in the amount of substructures from one system to another could also be responsible for part of the scatter of the relation. A similar bias should affect the measurement of the integrated Compton parameter $Y_X$, in case the average temperature is unaffected by the presence of substructure (see below).

\item
\emph{Entropy:} The entropy $K=Tn^{-2/3}$ is an important quantity to trace the state and the history of the ICM. In the presence of gas inhomogeneities, the measured entropy level could be significantly underestimated \citep{nagai}, which could be the reason for the relatively shallow entropy profiles observed in some systems \citep[and references therein]{walker13}. In addition, the temperature of the medium inside dense clumps is expected to be somewhat smaller than that of the underlying smooth ICM \citep{vazza12c}, which could bias the measured average temperatures low and enhance the flattening of the entropy profiles. A clumping factor of $\sim1.25$ at $R_{200}$, as inferred from this work, is sufficient to explain the relatively mild entropy slopes in cluster outskirts \citep{eckert13a} without the need for invoking additional mechanisms such as non-thermal pressure support \citep[e.g.,][]{lapi,kawa}; this result however needs to be confirmed with better precision. In addition, we note that biases in the spectroscopic X-ray temperatures \citep{mazzotta04,vikhlinin06b,leccardi07} would lower the observed entropy further and should be taken into account for the estimation of the entropy profiles.

\item
\emph{Hydrostatic mass:} A bias in the gas density would also affect the reconstruction of cluster masses through the hydrostatic equilibrium equation \citep[and references therein]{ettori13}. Indeed, the reconstructed mass profiles from X-ray data become

\begin{equation}
M_{\rm hyd}(<r)=-\frac{kT r}{\mu m_p G}\left(\frac{d\log T}{d\log r} + \frac{d\log n}{d\log r}+\frac{d\log\sqrt{C}}{d\log r}\right),
\end{equation}

and thus the observed hydrostatic masses are biased by the factor
\begin{equation}
\frac{M_{\rm obs}}{M_{\rm true}}=1+\frac{d\log\sqrt{C}}{d\log r}\left[\frac{d\log T}{d\log r} + \frac{d\log n}{d\log r}\right]^{-1}.
\end{equation}

\begin{figure}
\resizebox{\hsize}{!}{\includegraphics{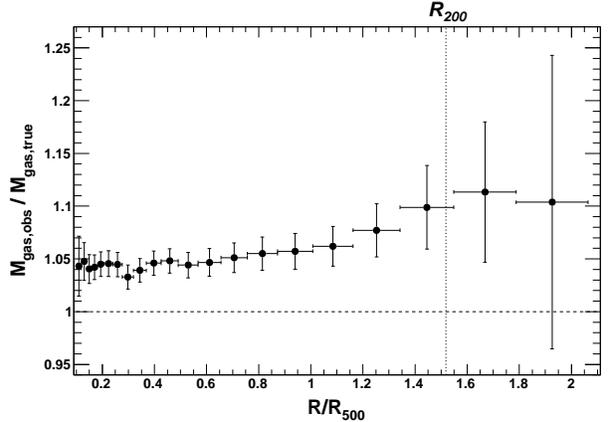}}
\caption{Bias in gas mass derived by comparing standard and clumping-corrected gas mass profiles (see Eq. \ref{eq:mgas}).\label{fig:mgas}}
\end{figure}

The reconstruction of hydrostatic mass profiles is thus affected only by the derivative of the clumping factor. As we can see in Fig. \ref{fig:clfprof}, the recovered clumping factors show only a mild dependence on radius, especially in comparison with the rather steep density gradients in cluster outskirts. At $R_{500}$, the typical values for the density and temperature gradients are $-2.0$ and $-0.5$, respectively, while for the emissivity bias we measure a gradient of $0.05$. Combining these values, we obtain a typical clumping bias of $\sim0.98$ at $R_{500}$. Thus, we can safely conclude that the effect of gas clumping on the measurements of cluster masses through X-ray data is negligible compared to other potential sources of systematics \citep[e.g.,][]{rasia06,nagai07,battaglia13b}. Conversely, it has been proposed to combine X-ray and Sunyaev-Zel'dovich probes to solve the hydrostatic equilibrium equation \citep{ameglio09,eckert13b}. In this case, since the density enters directly into the equation, hydrostatic masses are systematically underestimated by $\sqrt{C}$.

Recently, \citet{roncarelli13} analysed the effect of large-scale asymmetries on hydrostatic mass estimates in a sample of 62 clusters simulated with the Tree smoothed particle hydrodynamics (Tree-SPH) code \texttt{GADGET-3}, and reached the conclusion that the effect of clumping is in average negligible, in agreement with our observational results. However, they noted that the absolute value and the sign of the bias vary significantly from one system to another, which introduces a significant scatter in the scaling relations. This prediction could be tested in the future with higher-quality data.

\item
\emph{Gas fraction:} The gas fraction of galaxy clusters $f_{\rm gas}=M_{\rm gas}/M_{\rm hyd}$ is an important tool for cosmology \citep{white93,ettori03,allen08}, and a good understanding of the potential systematics affecting its measurements is important. The effect of clumping on the gas fraction is slightly enhanced with respect to the other quantities discussed here, since it combines the positive bias in gas mass with the negative bias in hydrostatic mass. At $R_{500}$, we find that $f_{\rm gas}$ is typically overestimated by a factor 1.08. This would cause an underestimation of $\Omega_m$ by the same factor when using the gas fraction to estimate the total matter content. The bias on $f_{\rm gas}$ introduced by clumping rises to 1.15 at $R_{200}$. This bias is significantly lower than the values inferred by \citet{simionescu} in the NW arm of Perseus.

\end{itemize}
\section{Discussion}
\label{sec:discussion}

\subsection{Comparison with previous works}

As shown in Fig. \ref{fig:clfprof}, our analysis indicates that the effect of clumping on X-ray observables is small out to $R_{500}$ and relatively mild out to $R_{200}$. A precise knowledge of this effect is important for our knowledge of the systematics in cluster scaling relations and the origin of the scatter of these relations. It also allows us to disentangle the effects of non-thermal pressure support from gas clumping. The results presented here agree with the recent results of high-sensitivity Sunyaev-Zeldovich (SZ) experiments \citep{planck5,bonamente12,sayers13}, which showed an excellent agreement between the pressure profiles obtained from X-rays and from the SZ effect inside $R_{500}$. Indeed, pressure profiles derived from X-rays would be significantly overestimated in the presence of strong inhomogeneities, while conversely the SZ effect is expected to be unaffected by clumping. Recent SZ results are thus confirming the very mild emissivity biases obtained here.

While the results presented here are the first obtained on a large cluster sample, a number of observational studies have been carried out to estimate the clumping factor and its radial dependence with different techniques. Recently, \citet{morandi13} proposed to relate the standard deviation of the surface brightness distribution to the clumping factor and tested this approach with a set of hydrodynamical simulations. This method was then applied to deep \emph{Chandra} observations of A1835 and A133 \citep{morandi13b}, which can probe clumping on smaller scales compared to our work. These studies imply a clumping factor $\sqrt{C}\sim1.5$ around $R_{200}$, which is slightly higher, but consistent with our results. We note that the method proposed by \citet{morandi13} requires the computation of a second-order quantity, which is more difficult to obtain observationally than the azimuthal median. Moreover, any additional surface-brightness fluctuations (caused, e.g., by the radial dependence of the surface brightness, cosmic variance, or Poisson noise) need to be carefully accounted for in this method.

Indirect methods have also been proposed to estimate the clumping factor. For instance, \citet{eckert13a} used the deviations of the observed entropy profiles to the self-similar expectation to infer the clumping factor in an indirect way. The recovered clumping factors in the outskirts are at the level of $\sqrt{C}\sim1.2-1.3$, in good agreement with the results presented here. A similar technique was recently applied by \citet{urban13} using \emph{Suzaku} observations of several radial arms of the Perseus cluster. The authors compared both the observed entropy and pressure profiles to the self-similar expectations and attributed the deviations of entropy to the density measurements rather than the temperature ones, as previously noted by \citet{e12}. This allowed \citet{urban13} to infer a gas clumping factor $\sqrt{C}\sim1.5$ at $R_{200}$ in Perseus. Therefore, a mild level of inhomogeneities in the gas distribution could be responsible for the rather flat entropy profiles observed in a few cases beyond $R_{500}$, as proposed originally by \citet{nagai}, without the need for invoking additional mechanisms such as non-thermal pressure support \citep[e.g.][]{fusco13}. A higher precision in the measurement of the clumping factor at $R_{200}$ is however required to test conclusively this hypothesis.

\subsection{Comparison with numerical simulations}

We compared the clumping-factor profiles obtained in this paper with several sets of hydrodynamical simulations to test the observed level of hot gas accretion with the one predicted in the $\Lambda$CDM paradigm. To compare between grid codes and smoothed particle hydrodynamics (SPH) simulations, we used the simulated cluster samples introduced in Sect. \ref{sec:sims}. We remind that they consist in a sample of 62 clusters and groups simulated using the Tree-SPH code \texttt{GADGET-3} (see Sect. \ref{sec:gadget}) and a set of 20 clusters simulated with the \texttt{ENZO} code (Sect. \ref{sec:enzo}). In the first case, we also investigated the effects of including baryonic physics in the simulations by comparing a non-radiative setup and a setup including gas cooling, star formation, galactic winds, and AGN feedback \citep[see][]{planelles13}. In all cases, the 3D clumping factor is estimated by computing directly $C=\langle n^2\rangle/\langle n\rangle^2$ in spherical shells. These profiles are then compared with the clumping factor profile presented in Fig. \ref{fig:clfprof} (right).

\begin{figure}
\resizebox{\hsize}{!}{\includegraphics{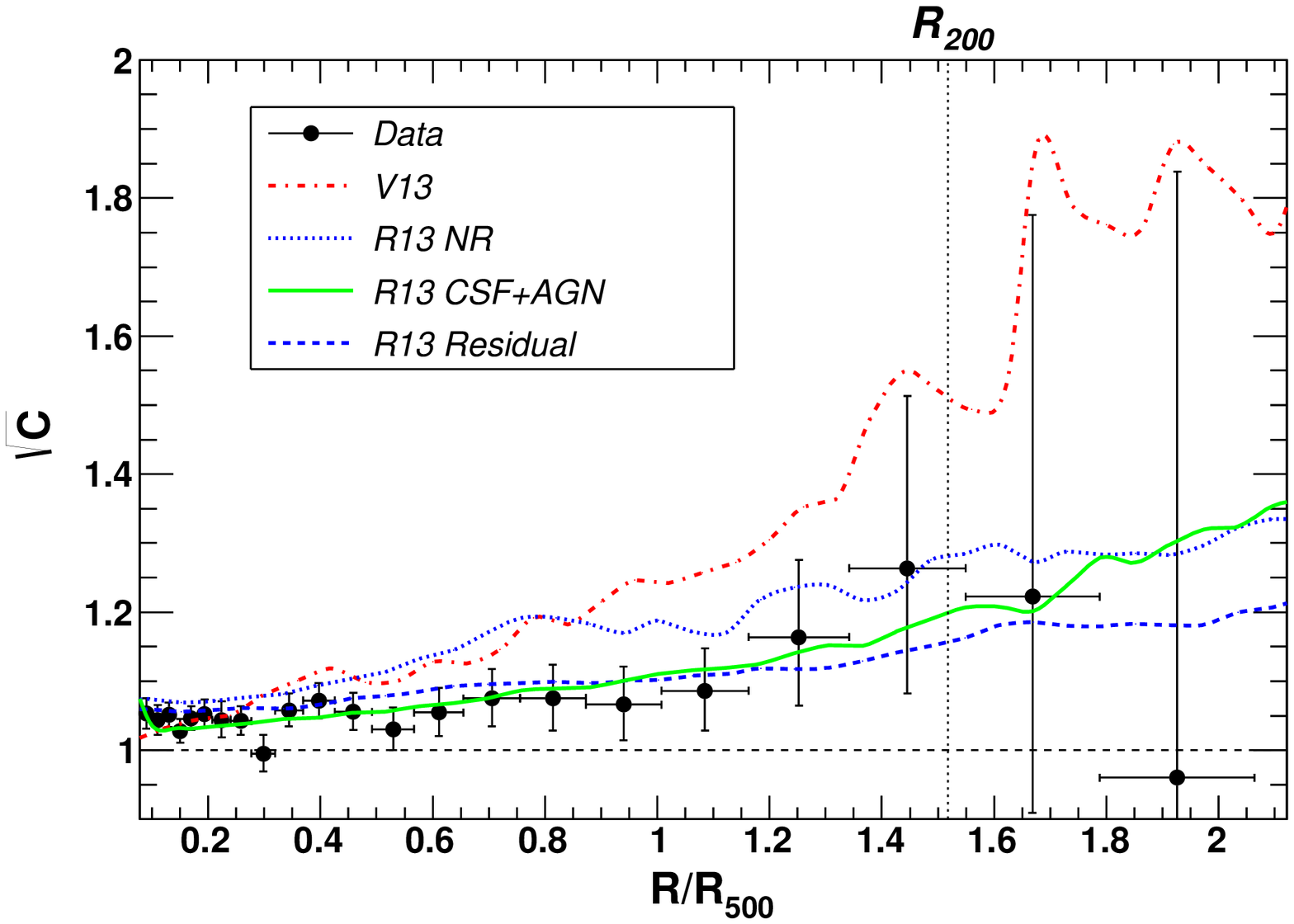}}
\caption{Comparison between the observed clumping factor (black data points) and the median 3D clumping factor profiles from three sets of numerical simulations: \texttt{ENZO} non-radiative \citep[dash-dotted red,][]{vazza12c}, \texttt{GADGET} \citep{roncarelli13} non-radiative (dotted blue) and including gas cooling, star formation, and AGN feedback (solid green). The dashed blue curve represents the ``residual clumping'' obtained after cutting out the densest parts of the distribution \citep{roncarelli13}.\label{fig:clfsim}}
\end{figure}

 As we can see in Fig. \ref{fig:clfsim}, out to $\sim1.2R_{500}$ the level of gas clumping predicted by the non-radiative simulations systematically overestimates the observed values, both in the grid and in the SPH case. At large radii, because of the large error bars we cannot provide any strong constraint on the models. Conversely, the Tree-SPH profile including radiative cooling, star formation, and AGN feedback \citet[green line]{roncarelli13} provides an excellent match to the observations. In this case, the lower clumping factor follows from the fact that cooling lowers the emission temperature of gas, therefore removing X-ray photons from the most structured component of the ICM \citep{fang09,lau11}. In addition, the cumulative effects of galactic winds and AGN feedback can contribute to smoothing the gas distribution within clusters. However, these latter runs tend to produce an amount of stellar mass which has some tension with observations \citep{planelles13}. We note that in a recent paper, \citet{battaglia14} showed that the volume used to compute the density and clumping factor profiles in SPH simulations could be incorrectly computed, which would affect the results presented here. However, this effect becomes important only beyond $R_{200}$, where our observational results are poorly constrained. At $R_{200}$ and inside, the proposed correction is only a few percent of the volume, thus the results presented here are qualitatively unaffected.
  
 In Fig. \ref{fig:clfsim} we also provide a comparison with the ``residual clumping'' in non-radiative SPH runs \citep{roncarelli13}, in which case the densest parts of the distribution were masked during the analysis (see Sect. \ref{sec:gadget}). Thus, only the effect of the inhomogeneities induced by large-scale accretion patterns is considered. This quantity was found to be robust against the included baryonic physics. The resulting profile (dashed blue) provides an excellent match to the data, which indicates that at first approximation, gas clumping in X-ray observations is mostly caused by the presence of large-scale inhomogeneities rather than compact clumps. 

It is worth noting that the comparison between observed and simulated datasets presented here depends on the choice of the sample, and in particular on the fraction of CC-like and NCC-like systems in all datasets. Indeed, accretion is likely more active in merging systems than in relaxed clusters, which could result in a larger clumping factor in the former class \citep[e.g.,][]{vazza12c}. In general, both simulated datasets are found to have a mix of ``perturbed" and ``relaxed" systems similar to the observed sample (see Sect. \ref{sec:sims}). In the \texttt{ENZO} sample this has been measured both by monitoring the matter accretion history of each system and through multi-pole decomposition of the X-ray surface brightness images and of the centroid shift of simulated cluster images \citep[][]{boehringer10}, finding that (with only one exception) the class of NCC-like clusters from the X-ray analysis overlaps with the class of actively merging objects, based on the mass accretion history, yielding  9 NCC-like objects and 11 CC-like objects. A similar X-ray analysis was also performed over \texttt{GADGET-3} clusters in \citet{roncarelli13}, resulting in the division of the simulated catalog into 30 and 32 objects, respectively. Based on that, we can assume that for most of the cluster volume our simulated samples present a ratio between CC and NCC objects that is similar to our observational dataset, and that since no big differences are found in the outer profiles of clusters around $R_{\rm 200}$ \citep{eckert13a} when CC and NCC clusters are compared, at first approximation we can just compare the average clumping factor of simulated and observed objects without further splitting the sample into dynamical classes.

Alternatively, a fraction of the true substructures may have been detected by our point-source detection procedure and masked during the analysis. This would lower the observed level of clumping and explain the excellent agreement with the profile obtained by masking the most prominent substructures in the non-radiative simulations (dashed blue line in Fig. \ref{fig:clfsim}). In this case, deep exposures with \emph{XMM-Newton} and \emph{Chandra} would be able to reveal these substructures and characterise them. Another possible explanation for the difference between non-radiative simulations and data could be the slight difference in resolution between the observed and simulated maps. Indeed, the mean resolution of our binned surface-brightness maps in the outskirts is of the order of $\sim100$ kpc (see Fig. \ref{fig:resolution}), which is about a factor of at least 4 larger than the resolution of the ENZO simulation runs. To investigate this, we convolved our simulated images with a Gaussian of 100 kpc radius, and performed the same analysis on the convolved images. The emissivity bias profile is only mildly affected by this procedure and remains significantly higher than the observed one, showing that the difference between non-radiative simulations and observations is unlikely to be caused by a resolution effect.

\section{Conclusion}
\label{sec:conclusion}

In this paper, we have presented a method based on the azimuthal median to recover unbiased gas density profiles from X-ray measurements in the presence of inhomogeneities. This method has been applied to a sample of 31 clusters observed with \emph{ROSAT}/PSPC \citep{e12} to infer the typical level of gas clumping in galaxy clusters. Our results can be summarized as follows:

\begin{itemize}
\item
Using extensive hydrodynamical and synthetic simulations, we have demonstrated that the azimuthal median is unbiased by the presence of inhomogeneities in the underlying gas distribution (see Sect. \ref{sec:validation}). From a set of clusters simulated with the Eulerian code \texttt{ENZO}, we have shown that this method is able to recover the true 3D density profiles with an accuracy better than 10\% at all radii. The typical bias induced by the $n_e^2$ dependence of the emissivity can thus be estimated through the quantity $b_X=S_{X,{\rm mean}}/S_{X,{\rm median}}$, the \emph{emissivity bias}. In addition, the clumping factor can be recovered by taking the ratio between the deprojected mean and median profiles (see Sect. \ref{sec:bx_vs_C}).  In practice, we proposed an efficient algorithm based on Voronoi tessellation to measure the azimuthal median from X-ray data (see Sect. \ref{sec:application}).

\item
Applying this method to our cluster sample, we derived radial emissivity bias and clumping factor profiles (see Fig. \ref{fig:clfprof}) which can be used to estimate the typical bias in survey data. The recovered clumping factors are low ($\sqrt{C}<1.1$ out to $R_{500}$) and show a mild trend of increasing bias with radius, although the error bars in the outskirts are too large to reach a definite conclusion. A dataset with higher statistics and resolution would be needed to improve these constraints significantly. We also split our cluster population into CC and NCC subsamples to look for differences between dynamical states. With the exception of the inner regions, where CC systems appear slightly more regular than NCC clusters, we do not find any significant difference between the two populations.

\item
We used our average clumping factor profiles to investigate the effect of the bias in several derived quantities (see Sect. \ref{sec:derived}). We find that the gas mass is typically biased by a factor of $1.06$ at $R_{500}$. The effect on hydrostatic mass is smaller, at the level of a few percent. Since it combines the effects on gas mass and hydrostatic mass, the effect on gas fraction is the largest, with values ranging from 8\% at $R_{500}$ to 15\% at $R_{200}$. These numbers are very similar to the values predicted by \citet{roncarelli13} after the removal of the brightest clumps.

\item
 Comparing our results with the predictions of three different sets of numerical simulations obtained using the grid code \texttt{ENZO} (non-radiative) and the Tree-SPH code \texttt{GADGET-3} (a non-radiative run and a run including gas cooling, star formation, stellar winds, and AGN feedback), we find that non-radiative numerical simulations systematically overpredict the effect of gas clumping on X-ray observables, both in SPH and grid codes. The absence of radiative cooling in these simulations could be responsible for this discrepancy, since radiative energy losses remove the densest parts of the gas from the X-ray emitting regime. For this reason, the runs including cooling and feedback provide a better description of the data than the non-radiative runs (see Fig. \ref{fig:clfsim}). The ``residual clumping'' profiles \citep{roncarelli13} also provide an excellent match to the data, indicating that the emissivity bias is mostly caused by large-scale inhomogeneities rather than compact accreting structures. Alternatively, a number of substructures may have been picked out by our source detection algorithm and filtered out, in which case long exposures with \emph{Chandra} and \emph{XMM-Newton} will be able to detect and characterise them.

\end{itemize}

\section*{Acknowledgements} We acknowledge Klaus Dolag, Stefano Borgani and the Dianoga team for kindly providing us the Dianoga dataset. We thank Daisuke Nagai, Susana Planelles, Elena Rasia and the anonymous referee for helpful comments on the paper. F.V. acknowledges the usage of computational resources under the CINECA-INAF 2008-2010 agreement. F.V. gratefully acknowledges C. Gheller and G. Brunetti for their contribution to produce the cluster catalog used in this work. 

\bibliographystyle{mn2e}
\bibliography{clumping}

\normalsize

\begin{appendix}
\section{Relations between emissivity bias, azimuthal scatter and clumping factor for log-normal distribution of the gas density and X-ray surface brightness}
\label{app:lognormal}

In this section, we derive the relations between emissivity bias, azimuthal scatter and clumping factor under the assumption that, at each radius, the distributions of the gas density ($n$) and surface brightness ($S$) are log-normal. We refer to previous work \citep{kawa07,kawa08,khedekar13} for an introduction on, and detailed analysis of, the log-normal distribution of the ICM properties.

A log-normal distribution of the variable $X$ with a mean $\mu$ and a standard deviation $\sigma$ has the following properties: 
\begin{itemize}
\item[(i)] the expectation value $E(X^n)$ for $X^n$ is equal to $\exp(n \mu +n^2 \sigma^2/2)$; 
\item[(ii)] the variance of $X$ is $E(X^2)-E(X)^2 = (\exp(\sigma^2)-1) \exp(2 \mu+\sigma^2)$; 
\item[(iii)] the median $X$ is $\exp(\mu)$.
\end{itemize}

The gas density clumping factor is defined as $C = \langle n^2\rangle/\langle n\rangle^2$. By applying the relation (i) on the log-normal distribution, we can then write \begin{equation}
C = \frac{ \exp(2 \mu+ 2 \sigma^2) }{\exp(2\mu+\sigma^2)} = \exp(\sigma^2),
\end{equation}
where $\mu$ and $\sigma$ are hereafter the mean and standard deviation of the log-normal distribution of the gas density at given radius.

\begin{figure}
\resizebox{\hsize}{!}{\includegraphics{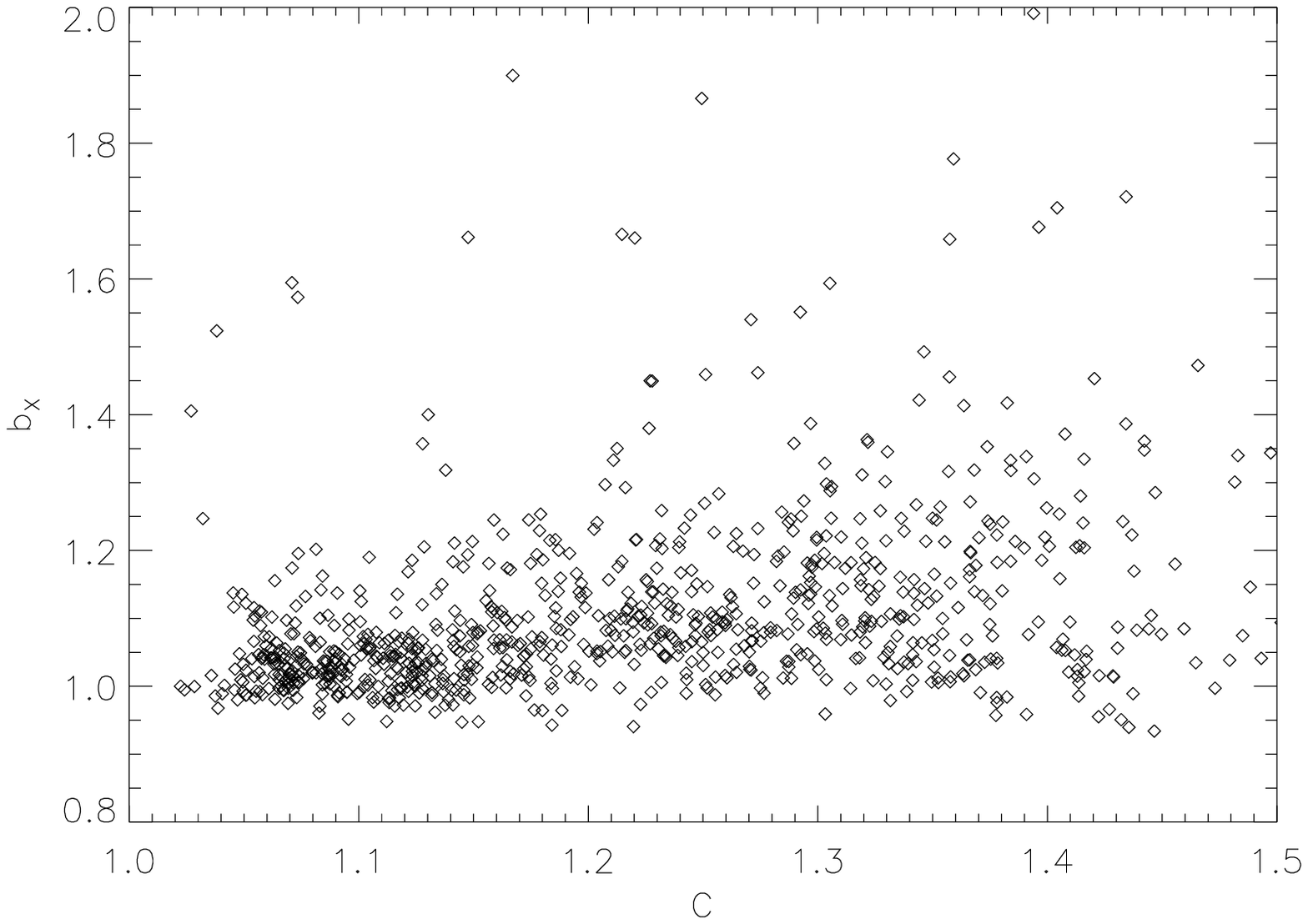}}
\caption{\label{fig:bx_C_sim}Relation between emissivity bias $b_X$ and clumping factor $C$ in the Dianoga sample \citep{roncarelli13}. }
\end{figure}

\citet{kawa08} discuss the implementation of a statistical method to extract the three-dimensional properties of the gas density fluctuations from the bi-dimensional X-ray surface brightness. To calibrate their method, they make use of synthetic galaxy clusters obtained by combining an isothermal $\beta-$model with the 3D power spectrum estimated from objects extracted from hydrodynamical cosmological simulations.
They conclude that the dispersion $\sigma_S$ of the log-normal distribution in each radial bin of the X-ray surface brightness $S$ is related to the dispersion $\sigma$ in the distribution of the gas density through the relation $\sigma_S = q \sigma$, where $q = c_1 / (c_2+|a_q|^{-4})$ and $a_q \approx a_S-0.2$, being $a_S$ the power-spectrum index of the fluctuations in the surface brightness map \citep[see equations 16-17, and sect.~3.2, in][]{kawa08}. The coefficients $c_1$ and $c_2$ calibrated for their synthetic objects are $2.05 \times 10^{-2}$ and $1.53 \times 10^{-2}$, respectively.
From our definition of the emissivity bias, $b_X = S_{mean}/S_{med}$, and using the relations (ii) and (iii) of the log-normal distribution, we can write: 
\begin{equation}
b_X = \exp(\sigma_S^2 / 2) = \exp(\sigma^2 q^2/2) = C^{q^2 / 2}.
\end{equation}
In Fig. \ref{fig:bx_C_sim} we show the relation between $b_X$ and $C$ in the Dianoga sample (see Sect. \ref{sec:bx_vs_C}). We indeed find a positive correlation between the two quantities (with a Spearman rank coefficient of 0.49), and the best-fit relation between the two quantities reads $b_X\sim C^{0.5}$. From this relation, we require $q=1$ and a power spectrum index $|a_S| \sim 2-2.5$, that falls into the range estimated by \citet{kawa08} from the hydrodynamical simulations of \citet{dolag05b} and then propagated to the synthetic objects.

Finally, let us consider the azimuthal scatter $s$.  From our definition in \citet{vazzascat},
\begin{equation} s^2 = \frac{1}{N} \sum_i^N \frac{ (S_i - \langle S\rangle)^2 }{\langle S\rangle^2},\end{equation}
where the sum is performed at each radius over the $N$ azimuthal sectors in which a surface brightness $S_i$ is estimated. Then, $\langle S\rangle^2 s^2$ is the variance of $S$ and, from the relation (ii) above, we can write, 
\begin{equation}\langle S\rangle^2 s^2 = (\exp(\sigma_S^2)-1) \exp(2\mu_S+\sigma_S^2),\end{equation}
where $\mu_S$ is the mean of the log-normal distribution of the surface brightness at given radius. The latter relation is strictly valid when $N$ is large enough to overcome any bias in the estimate of the sample variance.
Then using the relation (i), 
\begin{equation}
s^2 = \exp(\sigma_S^2)-1 = \exp(\sigma^2q^2)-1 = C^{q^2}-1 = b_X^2 -1,
\end{equation}
that implies $s \sim \sqrt{C -1}$ from our best-fit $b_X\sim C^{0.5}$.

\section{Surface brightness distribution in X-ray observations}
\label{app:sbdist}

The method presented in this paper relies on the assumption that the distributions of density within a radial shell and the distribution of X-ray surface brightness in an annulus are roughly log-normal. This assumption is supporter by numerical simulations \citep{zhuravleva13,khedekar13,rasia14}, but should be confirmed in actual data before the accuracy of the method presented here can be fully demonstrated. Unfortunately, because of the poor statistics and angular resolution of the \emph{ROSAT}/PSPC used in this paper, the data quality is insufficient to carry out such a study. However, with the appropriate strategy \emph{XMM-Newton}/EPIC data can be used for this purpose, thanks to the much larger effective area and the better PSF of this instrument. 

\begin{figure}
\resizebox{\hsize}{!}{\includegraphics{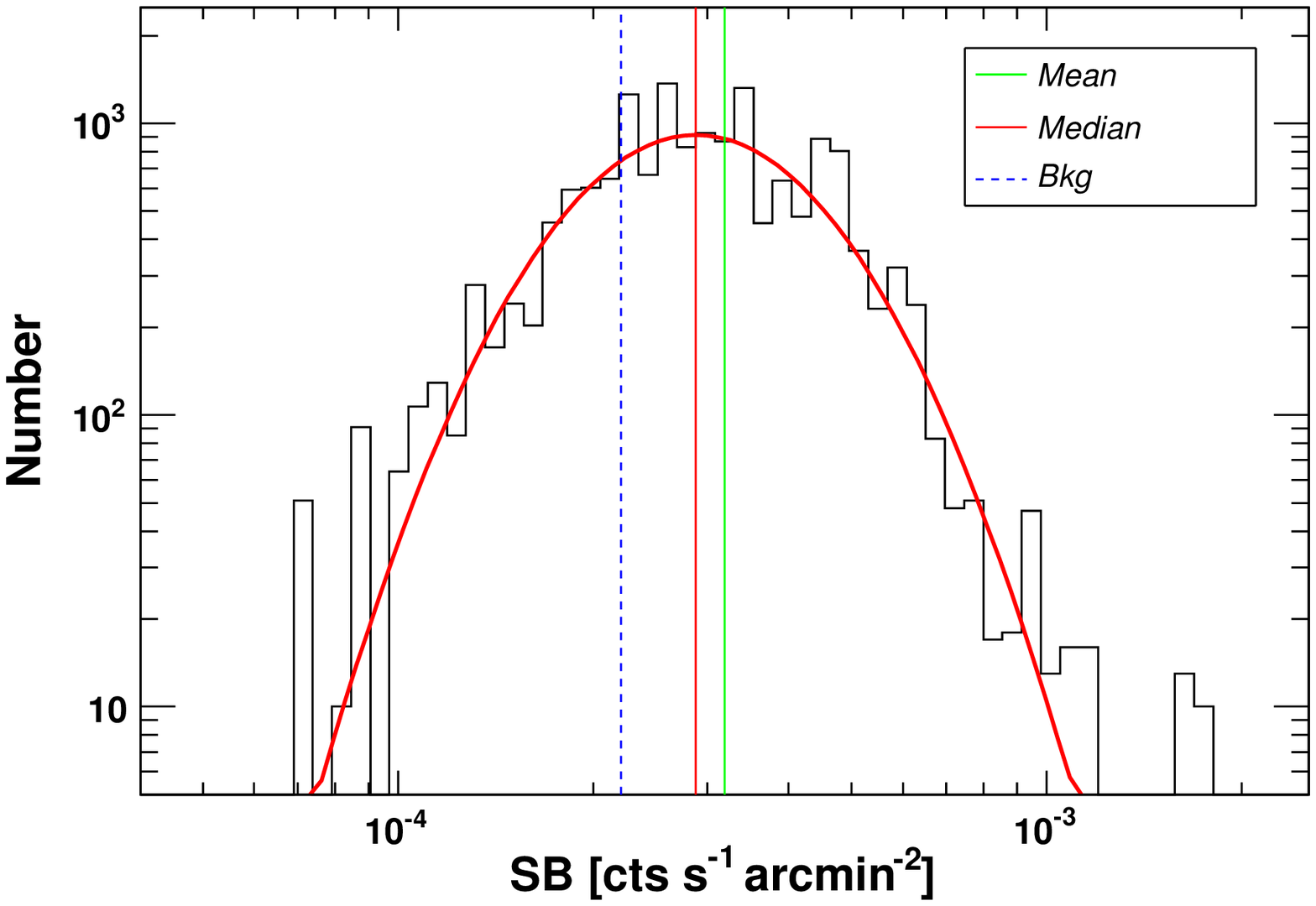}}
\caption{\label{fig:a2142} Distribution of surface brightness in the radial range 18-19 arcmin ($[0.83-0.87]R_{200}$) from an \emph{XMM-Newton} mosaic of the massive cluster A2142 (Eckert et al. in prep). The red curve shows the best-fit log-normal distribution, compared to the background expectation (dashed blue), the sample mean (green) and median (red).}
\end{figure}

We used a proprietary \emph{XMM-Newton} mosaic of the cluster A2142 \citep[proposal ID 069444,][]{eckert14b} to study the surface-brightness distribution of the cluster at $R_{500}$ and beyond. Voronoi tessellation was applied to the data (as explained in Sect. \ref{sec:application}) to generate a binned surface-brightness map, from which the distribution of pixel values in concentric annuli was extracted. In Fig. \ref{fig:a2142} we show the resulting distribution in the radial range 18-19 arcmin, which corresponds roughly to $[0.83-0.87]R_{200}$ for this cluster. We fitted the observed distribution with a log-normal distribution including the skewness as a free parameter, using the function

\begin{equation} \log N(x)= 2\phi(x)\Phi(\alpha x), \end{equation}

\noindent where $\phi(x)$ and $\Phi(x)$ represent the Gaussian probability density function and cumulative distribution, respectively, and $\alpha$ is the skewness parameter \citep{azzalini85}. The log-normal distribution is found to provide an excellent match to the observed distribution, with the sample mean and median significantly shifted from one another. The distribution is found to be symmetric, with a best-fit skewness of 0.03, and thus the median and the mode of the distribution are the same, whereas the mean is biased high by $\sim15\%$. This analysis thus demonstrates that the log-normal approximation is a valid representation of the surface-brightness (and density) distribution in cluster outskirts.

\end{appendix}
\end{document}